{}
{}

\documentclass[prd,reprint,superscriptaddress]{revtex4-1}

\newif\ifpublic\publictrue

\usepackage{fancyhdr}
\ifpublic\else
\fi
\newif\ifworking\workingtrue

\usepackage{mathtools,mathrsfs,amsbsy,amssymb,latexsym,amsfonts,amscd,
amsmath}
\usepackage{bm}
\usepackage{graphicx}
\usepackage[usenames,dvipsnames]{color}
\usepackage[normalem]{ulem}
\usepackage{natbib}
\definecolor{linkcolor}{rgb}{0,0,0.6}
\usepackage[ pdftex,colorlinks=true,
pdfstartview=FitV,
linkcolor= linkcolor,
citecolor= linkcolor,
urlcolor= linkcolor,
hyperindex=true,
hyperfigures=false]
{hyperref}

\def\showkeysrefformat#1{{\normalfont\tiny\ttfamily#1}}
\makeatletter
\def\SK@@ref#1>#2\SK@{%
{\@inlabelfalse\leavevmode\vbox to\z@{%
\vss\SK@refcolor\rlap{\vrule\raise .75em%
\hbox{\showkeysrefformat{#2}}}}}}
\makeatother

\allowdisplaybreaks[3]
 
\expandafter\def\expandafter\bfseries\expandafter{\bfseries\ifmmode\else\boldmath\fi}
\expandafter\def\expandafter\mdseries\expandafter{\mdseries\ifmmode\else\unboldmath\fi}
\expandafter\def\expandafter\normalfont\expandafter{\normalfont\ifmmode\else\unboldmath\fi}

\def\g{\mathfrak g}

\def\beq{\begin{equation}}
\def\eeq{\end{equation}}

\def\beqz{\begin{equation*}}
\def\eeqz{\end{equation*}}

\def\bea{\begin{eqnarray}}
\def\eea{\end{eqnarray}}

\def\ha{\mbox{\small $\frac{1}{2}$}}

\def\id{\protect{{1 \kern-.28em {\rm l}}}}

\newcommand{\ti}[1]{_{\bm{{#1}}}}
\def\1{{\ti{1}}}
\def\2{{\ti{2}}}
\def\3{{\ti{3}}}

\newcommand{\s}{\sigma}

\newcommand{\gb}[1]{g^{(#1)}}
\newcommand{\dd}{\text{d}}
\newcommand{\jb}[1]{j^{(#1)}}
\renewcommand{\k}[1]{\kay_{#1}}
\newcommand{\Ww}[1]{I_{\text{W}\hspace{-1pt}\text{Z}}\bigl[#1\bigr]}
\newcommand{\W}[1]{I_{\text{W}\hspace{-1pt}\text{Z}}\left[#1\right]}
\newcommand{\p}{\partial}

\newcommand{\ze}{\zeta}

\definecolor{brightBlue}{rgb}{0,0,1}

\definecolor{Violet}{rgb}{0.47,0,1}

\DeclareFontEncoding{LS1}{}{}
\DeclareFontSubstitution{LS1}{stix}{m}{n}
\DeclareSymbolFont{stixsymbols}{LS1}{stixscr}{m}{n}
\SetSymbolFont{stixsymbols}{bold}{LS1}{stixscr}{b}{n}
\DeclareMathSymbol{\kay}{\mathalpha}{stixsymbols}{"6B}

\ifpublic

\else

\newcounter{comcompt}
\newcounter{piqle}
\newcounter{treqle}
\newcounter{boxq}
\fi

\ifworking
\newcommand{\work}[1]{\begingroup \makeatletter\def\f@size{8}
#1 \endgroup}
\else
\newcommand{\work}[1]{\ignorespaces}
\fi
\newcommand{\workstep}[1]{\scriptsize $\bullet$~#1 \makeatletter\def\f@size{8}
}

\begin{document}
\hfill [ZMP-HH/18-26]

\bigskip

\title{Integrable coupled \texorpdfstring{$\sigma$-models}{sigma-models}
} 

 
\author{F. Delduc}
\affiliation{Univ Lyon, Ens de Lyon, Univ Claude Bernard, CNRS, Laboratoire de Physique, \\
F-69342 Lyon, France}
\author{S. Lacroix}
\affiliation{II. Institut f\"ur Theoretische Physik, Universit\"at Hamburg, 
\\
Luruper Chaussee 149, 22761 Hamburg, Germany \\
Zentrum f\"ur Mathematische Physik, Universit\"at Hamburg, \\
Bundesstrasse 55, 20146 Hamburg, Germany}
\author{M. Magro}
\affiliation{Univ Lyon, Ens de Lyon, Univ Claude Bernard, CNRS, Laboratoire de Physique, \\
F-69342 Lyon, France}
\author{B. Vicedo}
\affiliation{Department of Mathematics, University of York, \\
York YO10 5DD, UK}
\begin{abstract}
A systematic procedure for constructing classical integrable field theories with arbitrarily 
many free parameters is outlined. It is based on the recent interpretation of 
integrable field theories as realisations of affine Gaudin models. In this language, 
one can associate integrable field theories with affine Gaudin models having arbitrarily 
many sites. We present the result of applying this general procedure to couple together an arbitrary number of principal chiral model fields on the same Lie group, each with a Wess-Zumino term.
\end{abstract}
\pacs{x}
\maketitle

\section{Introduction}

The very scarceness of the property of integrability in classical and quantum systems makes its ubiquity in high energy physics as well as its rich history in condensed matter physics seem even more remarkable. For instance, integrability has played a pivotal role in recent years (see \emph{e.g.} the review \cite{Beisert:2010jr}) in the study of the various instances of the AdS/CFT correspondence, which brought about a new surge of interest in the field of both classical and quantum integrability. In particular, it has led to the development of numerous new methods as well as to the discovery of new integrable models.

One particular aspect of the subject of classical integrable field theories which has proved extremely fruitful in some of these recent developments is the concept \cite{Maillet:1985ec, Reyman:1988sf, Sevostyanov:1995hd, Vicedo:2010qd} of the twist function. This is a rational function $\varphi(z)$ of the spectral parameter $z$ which, for a very broad family of classical integrable field theories, controls the integrable structure of these theories through the Poisson bracket of the Lax matrix. It was proved \cite{Maillet:1985ec, Sevostyanov:1995hd, Vicedo:2010qd, Ke:2011zzb, Ke:2011zzc, Delduc:2013fga, Hollowood:2014rla, Hollowood:2014qma, Delduc:2014uaa, Vicedo:2015pna, Delduc:2015xdm, Schmidtt:2017ngw, Schmidtt:2018hop} that this family includes non-linear integrable $\sigma$-models and in particular ones which are relevant to the study of the AdS/CFT correspondence. In this setting, the twist function can be used,  for instance, to deform these theories while preserving integrability \cite{Delduc:2013fga, Vicedo:2015pna}.

It was recently shown by one of us in \cite{Vicedo:2017cge} that classical integrable field theories admitting a twist function can be seen as realisations of classical Gaudin models associated with an affine Kac-Moody algebra. Gaudin models are usually associated with finite dimensional Lie algebras, in which case they are thought of as spin chains. In the $\mathfrak{su}(2)$ case they can be obtained as limits of the inhomogeneous Heisenberg XXX chain. It may therefore seem somewhat surprising, at first, that an integrable field theory can be recast as an affine Gaudin model. However, it is important to stress that the spin chain in the affine case is not being used in the conventional way, as a discretisation of the field theory in which fields arise from taking a continuum limit by letting the number of sites go to infinity. Rather, classical fields emerge upon realising the underlying affine Kac-Moody algebra as a centrally extended current algebra on the circle.

Viewing the above class of integrable field theories in this new light opens up the possibility of constructing new classical integrable field theories by considering more general affine Gaudin models. Indeed, all the examples of integrable field theories discussed in \cite{Vicedo:2017cge} essentially correspond to affine Gaudin models with a single site. It is then natural to seek to construct field theories which correspond to affine Gaudin models with an arbitrary number of sites. The concrete implementation of such a construction will be presented elsewhere \cite{dlmvtoappear}, focusing on the class of field theories which are described by the simplest class of Gaudin models, namely the non-cyclotomic ones. Classical integrable field theories which are known~\cite{Vicedo:2017cge} to be realisations of this class of affine Gaudin models include the principal chiral model and integrable $\sigma$-models obtained from it by adding a Wess-Zumino term or by performing a non-abelian T-duality \cite{Fridling:1983ha,*Fradkin:1984ai}. Various integrable deformations of the principal chiral model also belong to this class, namely the inhomogenous \cite{Klimcik:2002zj, *Klimcik:2008eq} and homogeneous \cite{Kawaguchi:2014qwa} Yang-Baxter deformations as well as the $\lambda$-deformation \cite{Sfetsos:2013wia}. Finally, this class also contains the inhomogeneous Yang-Baxter deformation with a Wess-Zumino term constructed in \cite{Kawaguchi:2011mz, *Kawaguchi:2013gma} and \cite{Delduc:2014uaa}.
  
To illustrate the main idea behind the construction it is useful to recall that, when regarded as an integrable spin chain, a (classical) Gaudin model has the following two salient features:
\begin{itemize}
  \item[$(i)$] the degrees of freedom, \emph{i.e.} `spins', at different sites mutually (Poisson) commute, and
  \item[$(ii)$] the Hamiltonian describes interactions between the spins at every pair of sites.
\end{itemize}
Consider an affine Gaudin model with an arbitrary number of sites $N$. It follows from property $(i)$ that its phase space is a Cartesian product of $N$ separate phase spaces attached to individual sites. We can take the latter to be the phase spaces of $N$ different integrable field theories, each described by single site Gaudin models associated with the same affine Kac-Moody algebra. It then follows from property $(ii)$ that the $N$ integrable field theories which we chose to attach to the different sites will become coupled, with the strength of the couplings determined by the relative location of each site in the complex plane.

When the $N$ integrable field theories we start with are Lorentz invariant, it is natural to also require the coupled theory to be relativistic. There are certain necessary and sufficient conditions restricting the choice of Hamiltonian for the coupled theory which ensure that this is the case. We obtain, in this way, a relativistic integrable field theory which couples $N$ relativistic integrable field theories.

The purpose of the present letter is to report on the result of applying this procedure to $N$ coupled principal chiral model fields on the same Lie group, each with a Wess-Zumino term. In particular, we present the action of the resulting coupled $\sigma$-model and give the Lax connection underlying its integrability.

\section{The model}

\subsection{Action}

Let us fix a set of $3N + 1$ real parameters $z_1, \ldots, z_N$, $\zeta_1^\pm, \ldots, \zeta_N^\pm$ and $\ell^\infty$. It will be convenient to gather these together into a single rational function of the spectral parameter $z$, with double poles in the set ${\cal P}=\{z_1, \ldots, z_N\}$ and simple zeroes in $\cal Z^+ \cup \cal Z^-$ where ${\cal Z}^\pm=\{\zeta_1^\pm, \ldots, \zeta_N^\pm\}$. Explicitly, we define
\begin{equation} \label{factorised twist}
\begin{split}
\varphi(z) &= - \ell^\infty \varphi_+(z) \varphi_-(z),\\
\varphi_\pm(z) &= \frac{\prod_{r=1}^N(z-\zeta_r^\pm)}{\prod_{r=1}^N(z-z_r)}.
\end{split}
\end{equation}
In the language of affine Gaudin models, the case $N=1$ corresponds to the twist function of 
the principal chiral model with a Wess-Zumino term \cite{Vicedo:2017cge}.
For arbitrary $N$, the above rational function $\varphi(z)$ can be taken as the twist function for an affine Gaudin model with $N$ sites.

In terms of the above data, the action obtained in \cite{dlmvtoappear} from this $N$-site affine Gaudin model takes the form
\begin{multline}\label{Eq:Action}
S\bigl[ \gb 1, \ldots, \gb N \bigr] = \iint \dd t \, \dd x \; \sum_{r,s=1}^N \; \rho_{rs} \, \kappa\big(\jb r_+, \jb s_-\big)  \\
+ \sum_{r=1}^N \k r \; \Ww {\gb r},
\end{multline}
where $g^{(r)}$ for $r = 1, \ldots, N$ are fields valued in a real Lie group $G_0$. We define the currents $j_\pm^{(r)}= g^{(r)-1} 
\partial_\pm g^{(r)}$ and the light-cone derivatives are given by $\partial_\pm=\partial_t\pm\partial_x$. 
In this expression $\kappa$ is the opposite of the Killing form 
on $\mathfrak{g}_0$, the Lie algebra of $G_0$. $\W g$  is the Wess-Zumino term
\beqz
\W g = \iiint \dd t \,\dd x \, \dd \xi \; \kappa\Bigl( g^{-1} \p_\xi g, \bigl[ g^{-1} \p_x g, g^{-1} \p_t g \bigr] \Bigr).
\eeqz
The two dimensional domain with coordinates $(t,x)$ (which may be a cylinder or a plane) is the boundary of the three dimensional domain with coordinates $(t,x,\xi)$.
The coefficients $\rho_{rs}$ and $\kay_r$ are determined from the factorisation of the twist function in \eqref{factorised twist} as
\begin{align}
\rho_{rr} &= \frac{\ell^\infty}{4} \big( \varphi'_{+,r}(z_r) \varphi_{-,r}(z_r) - \varphi_{+,r}(z_r) 
\varphi'_{-,r}(z_r) \big), \nonumber\\
\rho_{rs} &= \frac{\ell^\infty}{2} \frac{\varphi_{+,r}(z_r) \varphi_{-,s}(z_s)}{z_r - z_s},\label{lagp}\\
\kay_r &= \frac{\ell^\infty}{2} \big( \varphi'_{+,r}(z_r) \varphi_{-,r}(z_r) + \varphi_{+,r}(z_r) \varphi'_{-,r}(z_r) \big),\nonumber
\end{align}
for any $r \neq s$. 
Here we introduced for each $r = 1, \ldots, N$ the two rational functions
\begin{equation*}
\varphi_{\pm, r}(z) = (z-z_r) \varphi_\pm(z),
\end{equation*}
which are regular at $z_r$.

We note that there is some redundancy in the choice of the $3N+1$ parameters encoded 
in the twist function \eqref{factorised twist}. Indeed, the expressions \eqref{lagp} for the 
coefficients entering the action are invariant under the transformation
\begin{equation*}
z_r \longmapsto a z_r + b, \qquad \ell^\infty \longmapsto a^{-1} \ell^\infty,
\end{equation*}
for any real constants $a$ and $b$, so that there are a total of $3N-1$ free parameters 
in the action \eqref{Eq:Action}, including the overall factor $\ell^\infty$. In the case 
when $N=1$ this gives two free parameters $\rho_{11}$ and $\kay_1$, as expected for the action of a single principal chiral model field $g^{(1)}$ with a Wess-Zumino term.

The action \eqref{Eq:Action} describes a $\s$-model on the target space $G_0^{\times N}$. The geometry of this target space restricted to the $r^{\rm{th}}$-copy of $G_0$ alone is given by a metric proportional to the Killing metric and controlled by the coefficient $\rho_{rr}$, along with a $B$-field originating from the Wess-Zumino term, controlled by the parameter $\k r$. This describes a self-interaction of the field $\gb r$ with itself, similar to the one of a single principal chiral model with Wess-Zumino term.
In addition to these self-interactions terms, the action \eqref{Eq:Action} also contains couplings between the different fields $\gb r$, controlled by the off-diagonal coefficients $\rho_{rs}$. As these coefficients are not symmetric, these interaction terms contribute to both the metric and the $B$-field of the target space geometry on $G_0^{\times N}$.

\subsection{Field equations and Lax connection}

The field equations obtained from extremising the action \eqref{Eq:Action} with respect to the $N$ fields $g^{(r)}(x,t)$, taking into account relations \eqref{lagp}, may be written as
\begin{align}
&\frac{\varphi'_{+,r}(z_r)}{\varphi_{+,r}(z_r)}\partial_+j_-^{(r)}-\frac{\varphi'_{-,r}(z_r)}{\varphi_{-,r}(z_r)}\partial_-j_+^{(r)} \notag\\
&\; +\sum_{s\neq r}\bigg(\frac{\varphi_{-,s}(z_s)}{\varphi_{-,r}(z_r)(z_r-z_s)} \big(\partial_+j_-^{(s)}+[j_+^{(r)},j_-^{(s)}] \big) \label{fe}\\
&\qquad- \frac{\varphi_{+,s}(z_s)}{\varphi_{+,r}(z_r)(z_r-z_s)} \big( \partial_-j_+^{(s)}+[j_-^{(r)},j_+^{(s)}] \big) \bigg)=0. \notag
\end{align}

The $\sigma$-model defined by the action \eqref{Eq:Action} is integrable by construction. The light-cone components of its Lax connection can be expressed succinctly, again in terms of the factorisation \eqref{factorised twist} of the twist function, as
\beq
\mathcal L_\pm(z, x, t) = \sum_{r=1}^N \frac{\varphi_{\pm,r}(z_r)}{\varphi_{\pm,r}(z)} 
j_\pm^{(r)}(x,t).
\eeq
Making use of the identity
\begin{equation} \label{eq phirs}
\frac{1}{\varphi_{\pm,s}(z_r)} = \delta_{r,s} \frac{1}{\varphi_{\pm,r}(z_r)},
\end{equation}
for $r, s = 1, \ldots, N$, we note that $\mathcal L_\pm(z_r, x, t)=j^{(r)}_\pm(x,t)$. The zero curvature equation for the Lax connection
\begin{equation} \label{LZ}
\partial_+\mathcal L_--\partial_-\mathcal L_++[\mathcal L_+,\mathcal L_-]=0
\end{equation}
holds for any value of the spectral parameter $z$. When multiplied by $\prod_{r=1}^N(z-\zeta_r^+)(z-\zeta_r^-)$, the left hand side of \eqref{LZ} becomes a polynomial in $z$ of degree $2N-1$. It therefore leads to $2N$ equations among the fields. The first $N$ of these, obtained by setting $z=z_r$, are the Maurer-Cartan equations for the currents $j_\pm^{(r)}(x,t)$. The remaining $N$ equations are most easily obtained by taking a derivative of \eqref{LZ} with respect to $z$ and then setting $z=z_r$. Making use of the identity \eqref{eq phirs} and of
\begin{equation*}
\frac{d}{dz}\left(\frac{\varphi_{\pm, r}(z_r)}{\varphi_{\pm, s}(z)}\right)\bigg|_{z=z_r}=\frac{1}{z_r-z_s},
\end{equation*}
one checks that the last $N$ equations coincide with \eqref{fe}.

\subsection{Twist function and integrability}

The integrability of the field theory defined by the action \eqref{Eq:Action}, that is the existence of an infinite family of integrals of motion in involution, is guaranteed from the outset since it was obtained as a realisation of an affine Gaudin model \cite{Vicedo:2017cge}. In particular, by construction the Lax matrix $\mathcal L = \ha (\mathcal L_+ - \mathcal L_-)$ has the following (equal time) Poisson bracket with itself which takes on the special $r/s$-form \cite{Maillet:1985fn, *Maillet:1985ek}
\begin{align*}
\{ \mathcal L_{\1}(z, x), \mathcal L_{\2}(w, y) \} &= [\mathcal R_{\1\2}(z,w), \mathcal L_{\1}(z, x)] \delta(x-y)\\
&- [\mathcal R_{\2\1}(w,z), \mathcal L_{\2}(w, y)] \delta(x-y)\\
&- \big( \mathcal R_{\1\2}(z,w) + \mathcal R_{\2\1}(w,z) \big) \delta'(x-y),
\end{align*}
where $\delta(x - y)$ is the Dirac $\delta$-distribution. Here we have defined the $\mathcal R$-matrix
\begin{equation*}
\mathcal R_{\1\2}(z,w) = \frac{C_{\1\2}}{w - z} \varphi(w)^{-1}
\end{equation*}
where $C_{\1\2}$ is the split Casimir of the complexification $\g$ of $\g_0$. The twist function $\varphi(z)$ is the same as in \eqref{factorised twist}.

The twist function is known to play a fundamental role in many aspects of the classical integrability of field theories which can be realised as affine Gaudin models. For instance, in these theories, an infinite family of Poisson commuting local charges can be constructed as the integrals of certain invariant polynomials of $\varphi(z) \mathcal L(z, x)$ evaluated at the zeroes of the twist function \cite{Lacroix:2017isl}.

\subsection{Decoupling limit}

In the model just described, a given field $g^{(r)}(x,t)$ at site $z_r$ is coupled to all the other fields $g^{(s)}(x,t)$ at sites $z_s$. It is possible to find a limit in which two subsystems are pulled apart at infinite distance, and cease to interact. The limit will be obtained by letting a real parameter $\gamma$ tend to zero. We consider two models identical to the one described in the previous paragraph, with respectively $N_1$ and $N_2$ sites such that $N_1+N_2=N$ and with parameters $z^{(a)}_r$, $\zeta^{(a)\pm}_r$, $r = 1, \ldots, N_a$ for $a=1,2$. The latter determine factorisations of the twist functions $\varphi^{(a)}$, exactly as in \eqref{factorised twist},
\begin{align*}
\varphi^{(a)}(z) &= - \ell^\infty \varphi^{(a)}_+(z) \varphi^{(a)}_-(z),\\
\varphi^{(a)}_\pm(z) &= \frac{\prod_{r=1}^{N_a}\big( z-\zeta_r^{(a)\pm} \big)}{\prod_{r=1}^{N_a} \big( z-z^{(a)}_r \big)}.
\end{align*} 
The parameters of the whole system are related to those of the two subsystems by
\begin{align*}
1\leq r\leq N_1, \quad &z_r=z^{(1)}_r, \zeta^{\pm}_r=\zeta^{(1)\pm}_r,\\
1\leq r\leq N_2, \quad &z_{N_1+r}=z^{(2)}_r+ \gamma^{-1},\,\zeta^{\pm}_{N_1+r}=\zeta^{(2)\pm}_r+ \gamma^{-1}.
\end{align*}
Equivalently, we can write this more succinctly as
\begin{equation*}
\varphi_\pm(z)=\varphi^{(1)}_\pm(z)\varphi^{(2)}_\pm(z - \gamma^{-1}).
\end{equation*}
Notice that one might have applied any permutation to the sets $\cal P$ and $\cal Z^\pm$ before applying this procedure, so that one may choose freely the poles and zeroes of the two models which are going to decouple. Then one checks explicitely that in the decoupling limit when $\gamma$ tends to zero one has
\begin{align*}
S[g^{(1)},\ldots, g^{(N)}]\longrightarrow &\; S^{(1)}[g^{(1)},\ldots, g^{(N_1)}]\\
&\qquad + S^{(2)}[g^{(N_1+1)},\ldots, g^{(N)}],
\end{align*}
where $S^{(a)}$ for $a=1,2$ is the action of the subsystem with $N_a$ sites. We also have the limits
\begin{equation*}
\varphi_\pm(z) \longrightarrow \varphi^{(1)}_\pm(z), \qquad
\varphi_\pm(z+ \gamma^{-1})\longrightarrow \varphi^{(2)}_\pm(z),
\end{equation*}
as $\gamma \to 0$, together with
\begin{align*}
\mathcal L_\pm(z, x, t) &\longrightarrow \mathcal L^{(1)}_\pm(z, x, t),\\
\mathcal L_\pm(z+\gamma^{-1}, x, t) &\longrightarrow \mathcal L^{(2)}_\pm(z, x, t),
\end{align*}
where $\mathcal L^{(a)}_\pm(z, x, t), a=1,2$, is the Lax connection of the subsystem with $N_a$ sites. Thus one sees that in  the decoupled limit where the system is composed of two uninteracting subsystems, the complete Lax connection is naturally valued in $\g \oplus \g$.

\subsection{Global symmetries}

Since the action \eqref{Eq:Action} only depends on the left invariant currents $j^{(r)}_\pm$, the model with $N$ sites is invariant under the action of $G_0^{\times N}$ given by
\begin{equation*}
(g^{(1)},\ldots,g^{(N)}) \longmapsto (h_1g^{(1)},\ldots, h_N g^{(N)}), \quad h_r\in G_0.
\end{equation*}
The corresponding Noether currents are most simply obtained by conjugating the field equation \eqref{fe} by $g^{(r)}$ and rewriting it as $\partial_+ \mathcal K^{(r)}_- + \partial_- \mathcal K^{(r)}_+ = 0$ where
\begin{align*}
\mathcal K^{(r)}_\pm &= \pm \frac{\varphi'_{\pm,r}(z_r)}{\varphi_{\pm,r}(z_r)} g^{(r)}j_\pm^{(r)}g^{(r)-1}\\
&\qquad\qquad \mp \sum_{s \neq r} \frac{\varphi_{\pm,s}(z_s)}{\varphi_{\pm,r}(z_r)(z_r-z_s)}g^{(r)}j_\pm^{(s)}g^{(r)-1}.
\end{align*}

Because the currents at the various sites are all coupled, the symmetry under right multiplication is limited to the diagonal action of $G_0$ given by
\begin{equation*}
(g^{(1)},\ldots,g^{(N)}) \longmapsto (g^{(1)} h,\ldots, g^{(N)} h), \quad h\in G_0.
\end{equation*}
The corresponding Noether current reads
\begin{equation*}
\mathcal K_\pm = \pm \sum_{r=1}^N \varphi_{\pm, r}(z_r) j^{(r)}_\pm.
\end{equation*}
Its conservation $\partial_+ \mathcal K_- + \partial_- \mathcal K_+ = 0$ can be obtained by taking a linear combination of the $N$ equations of motion \eqref{fe} to eliminate the commutator terms.
This symmetry is enhanced in the decoupling limit, described in the previous paragraph, to $G_0 \times G_0$.

\subsection{Examples}

As a simple illustration of the general action in \eqref{Eq:Action} let us consider 
the case of two sites, namely $N=2$, with the parameters entering 
the twist function \eqref{factorised twist} chosen as
\begin{equation*}
z_1=- z_2=\gamma^{-1}, \quad \ze^\pm_1 = \pm 1 +\gamma^{-1}, \quad 
\ze^\pm_2 = \pm 1 - \gamma^{-1}.
\end{equation*}
For the coefficients \eqref{lagp} in the action we then find
\begin{gather*}
\rho_{11} = \rho_{22} =  \frac{\ell^\infty}{4} ( 2-\gamma^2), \qquad
\kay_1 = - \kay_2 = - \frac{\ell^\infty}{8}  \gamma^3, \\
\rho_{12} =- \frac{\ell^\infty}{16} \gamma(\gamma-2)^2, \qquad
\rho_{21} =  \frac{\ell^\infty}{16} \gamma(\gamma+2)^2.
\end{gather*}
The decoupling limit $\gamma \to 0$ is given by
\begin{equation}
\rho_{11}, \rho_{22} \longrightarrow \frac{\ell^\infty}{2}, 
\qquad \rho_{12}, \rho_{21}, \kay_1, \kay_2 \longrightarrow 0. \label{decu-example}
\end{equation}
In this case the action \eqref{Eq:Action} therefore describes, for $\gamma \neq 0$, 
a coupling of two principal chiral fields $g^{(1)}$ and $g^{(2)}$ valued 
in the same real Lie group $G_0$. We note  the presence of 
a WZ term for each of these fields in the coupled theory.

It is also possible to couple two principal chiral fields without 
introducing WZ terms and with the same decoupling
limit \eqref{decu-example} by choosing the following set of parameters of 
the twist function \eqref{factorised twist}:
\begin{gather*}
z_1=-z_2 = \gamma^{-1}, \quad 
\ze^\pm_1 = \pm \sqrt{1+\gamma^{-2} +\sqrt{1+ 4 \gamma^{-2}}}, \\
\ze^\pm_2 = \mp \sqrt{1+\gamma^{-2} -\sqrt{1+4 \gamma^{-2}}}.
\end{gather*}

\section{Conclusion}  

The general procedure illustrated in this letter opens up the possibility of constructing infinite families of relativistic
integrable field theories with a given underlying Lie algebra. The basic building blocks for the construction are relativistic integrable field theories which can be realised as affine Gaudin models. An arbitrary number of these building blocks can then be assembled together to form a new relativistic integrable field theory. In turn, since the latter is an affine Gaudin model by definition, it could be used as a new building block for subsequent constructions.
Determining the scope of all possibilities certainly requires a thorough investigation.

The fact that the twist function of the model we have considered is a generic rational function with $N$ double poles and $2N$ simple zeroes makes it possible to see the importance of this object permeating through to the very expression of the action. Indeed, it is quite remarkable that the parameters of the action can be so conveniently encoded in a choice of factorisation of the twist function.

To illustrate the method, we have focused on the class of non-cyclotomic affine Gaudin models. Working within this class, it is possible to couple together any integrable $\sigma$-model from the list given in the introduction, provided they all involve the same Lie group. For instance, as a different example to the one presented in this letter, one could couple together $N-1$ principal chiral models, each with a Wess-Zumino term, and one homogeneous Yang-Baxter deformation of the principal chiral model. The resulting integrable $\sigma$-model will be presented in detail elsewhere \cite{dlmvtoappear}. We expect that the class of non-cyclotomic affine Gaudin models also includes the model constructed very recently in \cite{Georgiou:2018hpd} which couples together two $\lambda$-models.

Another natural direction for further study would be to extend the procedure to the class of cyclotomic affine Gaudin models, which includes \cite{Vicedo:2017cge} the symmetric space $\sigma$-model. The cyclotomic nature of the underlying affine Gaudin model is related to the $\mathbb{Z}_2$-grading associated with this $\sigma$-model, or in other words to its gauge symmetry. There are a number of interesting questions pertaining to this class of affine Gaudin models. For instance, it is not immediately clear what the gauge symmetry of an integrable field theory corresponding to a cyclotomic affine Gaudin model with multiple sites should be. Indeed, recall that the principal chiral model on a real Lie group $G_0$ has both a left and right $G_0$-symmetry. We may choose to gauge, in an equivalent way, any subgroup of $G_0$ either from the left or from the right. Choosing this subgroup $H$ in such a way that the coset, say on the right, $G_0/H$, is symmetric ensures that the resulting $\sigma$-model will be integrable. For the model presented in this letter, the situation is more complicated since we have shown that the left and right symmetries are different.  From the analysis in \cite{Vicedo:2017cge}, and in particular the fact that the gauge symmetry is related to the site at infinity (which is a fixed point of the action of the cyclic group), we expect to be able to construct a $G_0^{\times N}/H_{\rm diag}$ $\sigma$-model which is integrable. This situation would be reminiscent of the models considered in \cite{Guadagnini:1987ty} (see also \cite{PandoZayas:2000he}) when the two Lie groups $G$ and $G'$ there are the same. Another important question that will require a full analysis is the issue of the decoupling limit within this class.

A possible next step would be to generalise the whole procedure to the supersymmetric case. This would first require introducing the notion of classical affine Gaudin model associated with a Kac-Moody superalgebra. An obvious question in this setting is to look for type IIB superstring $\sigma$-models which could be realised as supersymmetric affine Gaudin models with multiple sites.

Having constructed classical integrable field theories with arbitrarily many parameters, a very important and natural question concerns the study of these models at the quantum level. It would be very interesting to determine the renormalisation group flow in these multi-parameter field theories and identify their fixed points. The recent results obtained for the models mentioned above in \cite{Georgiou:2018hpd} indicate a very rich structure in the renormalisation group flow of these models already in the case of two sites.
Moreover, the fact that the affine Gaudin model approach to integrable field theories enables one to naturally construct infinitely many new relativistic classical integrable field theories corroborates the idea that this approach may also be fruitful at the quantum level, even in the single site case 
\cite{Vicedo:2017cge, Lacroix:2018fhf, Lacroix:2018itd, Lacroix:2018njs}.

\medskip

\paragraph*{Acknowledgments.} We thank H. Samtleben for useful discussions
and G. Arutyunov and A. Tseytlin for useful discussions and comments on the draft. 
This work is partially supported by the French Agence Nationale de la Recherche (ANR) under grant
 ANR-15-CE31-0006 DefIS. 


%

\end{document}